# Evasion Attacks: How Adversarial Noise Bypasses ML Classifiers

Parker Hummel
Department of Math, Data & Technology
Minot State University
Minot, ND, 58707
parker.hummel@minotstateu.edu

Ryne Skabo
Department of Math, Data & Technology
Minot State University
Minot, ND, 58707
ryne.skabo@minotstateu.edu

Muhammad Abusaqer
Department of Math, Data & Technology
Minot State University
Minot, ND, 58707
muhammad.abusaqer@minotstateu.edu

Authors' version. Presented at the 58th Midwest Instruction and Computing Symposium (MICS 2026), Eau Claire, WI, March 27 to 28, 2026. Trustworthy Language Intelligence Lab, Minot State University.

## Abstract

This paper presents a reproducible, educational study of evasion attacks in image classification and text classification. A compact convolutional network trained on MNIST reached 98.63% clean test accuracy and was evaluated under two white-box attacks. Under FGSM, accuracy fell to 60.20% at $\varepsilon = 0.15$ and 1.72% at $\varepsilon = 0.30$; under PGD it fell to 32.47% and 0.41%, and a bit-depth-reduction defense recovered only part of the loss. In the second experiment, DistilBERT fine-tuned on the SMS Spam Collection reached 98.75% accuracy and a 94.96% F1-score, but a controlled sequence of pre-defined perturbations (character substitutions, whitespace noise, and a benign suffix) produced only modest probability shifts in most displayed examples and no flip from spam to ham. Adversarial vulnerability is strongly modality-dependent: the MNIST experiment is a clear evasion demonstration, whereas the text experiment is a controlled robustness evaluation. Robustness must be tested empirically rather than inferred from clean accuracy.

# 1. Introduction

Adversarial examples are intentionally perturbed inputs designed to alter a machine-learning model's prediction while remaining close to the original input under a chosen perturbation constraint [2], [16]. In the security literature, such manipulations are commonly studied under the broader category of evasion attacks, where the model has already been trained and the adversary modifies inputs at inference time rather than corrupting the training data [2]–[4]. This distinction is important because high clean-data accuracy does not, by itself, guarantee reliable behavior once inputs are exposed to malicious or carefully crafted perturbations [2], [3].

This problem matters because machine-learning systems are increasingly used in settings where erroneous predictions can have practical consequences, including content filtering, fraud detection, medical-support systems, and other automated decision pipelines. Prior work has shown that even highly accurate neural networks can become unstable under small, structured perturbations, especially in white-box settings where gradient information is available to the attacker [2]–[4]. In text classification, robustness questions are further complicated by the discrete nature of language, the need to preserve meaning, and the interaction between perturbations and tokenization [13]–[15].

The present paper studies this issue through two complementary experiments. The first experiment uses the MNIST handwritten-digit benchmark [1] and evaluates two canonical white-box image attacks: the Fast Gradient Sign Method (FGSM) [2] and Projected Gradient Descent (PGD) [3]. This image-based setting provides a clear and visually interpretable demonstration of adversarial vulnerability. The second experiment examines a text-classification setting by fine-tuning DistilBERT [8] on the SMS Spam Collection [9] using the Hugging Face Transformers and Datasets ecosystem [10], [11]. It then evaluates how the classifier's spam probability changes under a controlled sequence of small, pre-defined text perturbations and a simple normalization step.

Accordingly, this paper makes three practical contributions. It provides a reproducible image-based evasion demonstration using FGSM and PGD on MNIST; it presents a controlled, ethics-aware robustness evaluation of a DistilBERT spam classifier under mild text perturbations; and it offers a side-by-side comparison showing that robustness behavior is modality-dependent and must be established empirically rather than inferred from clean-data performance alone [2]–[4], [13]–[15]. The study does not propose a new attack method.

# 2. Related Work

Prior research on adversarial machine learning has developed a substantial body of work on image-based evasion, robustness evaluation, defenses, and more recently adversarial behavior in natural-language processing. Because the present paper combines one image experiment and one text experiment, the related-work discussion is organized around those same themes.

## 2.1 Image Evasion Attacks

The modern literature on adversarial examples is commonly traced to the finding that small, deliberately chosen perturbations can cause high-confidence neural-network misclassifications [16]. Goodfellow, Shlens, and Szegedy formalized this observation through FGSM, a one-step attack that uses the sign of the input gradient to produce an adversarial example efficiently [2]. FGSM remains an important baseline because it provides an interpretable and computationally light first-order attack, making it especially suitable for pedagogical and benchmark settings such as MNIST [1], [2].

Madry et al. extended this line of work by framing adversarial robustness as a min-max optimization problem and by popularizing PGD as a stronger iterative first-order attack baseline [3]. In practice, PGD is widely used because repeated gradient updates within an allowed perturbation region typically expose more severe weaknesses than single-step attacks [3]. For this reason, studies that compare FGSM and PGD often interpret PGD as the stronger white-box reference point when evaluating robustness [3], [4].

## 2.2 Text Robustness and Adversarial NLP

Adversarial NLP differs from image-based adversarial learning because language is discrete, semantically constrained, and highly sensitive to tokenization. HotFlip introduced a gradient-based approach for generating character- and word-level text perturbations, demonstrating that small edits can change text-classification predictions [13]. Subsequent work expanded this direction with stronger attack baselines and benchmarking frameworks. Jin et al. showed that BERT-based models can still be vulnerable under carefully designed natural-language attacks, and TextAttack later provided a reusable framework for text attacks, augmentation, and adversarial training [14], [15]. Together, these works show that robustness evaluation in NLP requires attention not only to label changes, but also to meaning preservation, fluency, and perturbation validity [13]–[15].

The present text experiment does not implement a full automated NLP attack pipeline. Instead, it adopts a controlled robustness design based on pre-defined, surface-level perturbations. This narrower scope is intentionally aligned with the project's educational goals. It allows the paper to study confidence shifts in a local DistilBERT classifier without moving into automated or deployment-oriented attack generation [8], [14], [15].

BERT established the pretrained-transformer paradigm for NLP [7], DistilBERT preserved much of its performance at lower cost [8], and comparative studies by Abusaqer and collaborators have shown that transformer models are strong baselines in text classification, including social-media harm detection [5], [6].

## 2.3 Robustness Evaluation and Defenses

A parallel line of work has emphasized that robustness claims must be evaluated carefully. Carlini and Wagner showed that many apparent defenses fail under stronger adaptive or more thoroughly tuned attacks, highlighting the risk of overstating security based on limited testing [4].

Robustness-oriented work has therefore emphasized stronger evaluation practices, adversarial training, and careful reporting of attack assumptions [3], [4]. In the context of the present paper, this literature motivates two design choices: first, the inclusion of both FGSM and PGD rather than relying on a single attack, and second, the cautious interpretation of simple preprocessing defenses as partial mitigations rather than conclusive protection. Unlike work that develops stronger attacks or defenses, or evaluates them at scale under specialized assumptions [2]–[4], [13]–[15], the present study is deliberately narrow: a reproducible, student-accessible comparison of a benchmark image evasion demonstration with a controlled text robustness study, whose contribution lies in the comparison and its framing rather than in methodological novelty.

# 3. Experimental Setup and Evaluation Protocol

This study evaluates adversarial robustness in two complementary classification settings: image classification and text classification. This dual-modality design was intentionally chosen to contrast a benchmark setting in which adversarial vulnerability is well established and visually interpretable (MNIST image classification) with a second setting in which robust behavior can be examined for a compact transformer-based text classifier. The first experiment uses MNIST with gradient-based image attacks [1]–[3]; the second uses DistilBERT on a public SMS spam dataset [7]–[11]. Both experiments were implemented as standalone Jupyter notebooks with fixed random seeds, explicit hyperparameter reporting, and automatically saved plots and tables. This reporting style is consistent with recent text-classification studies by Abusaqer and collaborators, which emphasize transparent documentation of datasets, model settings, and evaluation metrics [5], [6].

## 3.1 Experiment 1: Image-Based Evasion Attacks on MNIST

Experiment 1 focuses on image-based evasion using the MNIST handwritten-digit benchmark, which contains 60,000 training images and 10,000 test images across ten classes [1]. Each image was converted to a tensor and scaled to the range [0, 1], and the data were loaded with a batch size of 128. The target classifier was a compact convolutional neural network composed of two convolutional layers, a max-pooling layer, a fully connected hidden layer with 128 units, and a final ten-class output layer. The model was trained with cross-entropy loss and the Adam optimizer using a learning rate of $10^{-3}$ for two epochs, after which the learned weights were saved for reuse in later runs.

| Experiment | Dataset / Model | Key Settings | Primary Outputs |
|---|---|---|---|
| Image evasion (MNIST) | MNIST + compact CNN | Batch size = 128; Adam; learning rate = $10^{-3}$; 2 epochs; FGSM and PGD evaluated over $\varepsilon \in \{0.00, 0.02, 0.05, 0.10, 0.15, 0.20, 0.30\}$; PGD uses $\alpha = 0.01$ and 20 steps | Clean accuracy; FGSM / PGD accuracy-vs-ε curves; clean/adversarial/perturbation panel; simple defense comparison |
| Text robustness (SMS spam) | SMS Spam Collection + DistilBERT | 80/10/10 split; learning rate = $2 \times 10^{-5}$; batch sizes = 16/32; 1 epoch; pre-defined perturbation sequence plus normalization | Accuracy, precision, recall, F1; confusion matrix; probability trajectories under perturbation; normalization-effect plot |

Table 1: Summary of the experimental configuration and the primary outputs generated in the two experiments.

Two standard white-box attacks were evaluated on the MNIST classifier: the Fast Gradient Sign Method (FGSM) and Projected Gradient Descent (PGD). FGSM was selected because it provides a simple and widely used one-step gradient baseline for adversarial example generation [2]. PGD was included because iterative projected attacks are commonly treated as stronger first-order robustness evaluations than single-step attacks [3]. In both cases, adversarial examples were constrained to the valid image range [0, 1]. The FGSM and PGD experiments used the same perturbation-budget sweep, $\varepsilon \in \{0.00, 0.02, 0.05, 0.10, 0.15, 0.20, 0.30\}$. For PGD, the attack step size was $\alpha = 0.01$, the number of attack iterations was 20, and random initialization inside the allowable $L\infty$ perturbation region was enabled, with the random start seeded (seed 42) before each ε so that every run is reproducible.

The MNIST evaluation protocol included four parts. First, clean test accuracy was measured on unmodified images. Second, attacked test accuracy was computed for each ε value under both FGSM and PGD. Third, representative qualitative examples were saved to compare a clean image, its adversarial counterpart, and the corresponding absolute perturbation heatmap. Fourth, a simple preprocessing-based defense baseline was included through bit-depth reduction. In this defense setting, the same PGD adversarial inputs generated at $\varepsilon = 0.15$ for Table 2 were quantized to 16 intensity levels before classification. This defense was included only as an illustrative baseline rather than as evidence of strong protection, since prior work has shown that many apparently successful defenses do not remain effective under stronger evaluation [4].

To make the setup intuitive, the image experiment first performs a visual sanity check on randomly sampled clean digits before any attack is applied. Figure 1 shows representative clean MNIST examples together with their true labels and initial predictions; these examples help verify that the classifier behaves sensibly on unmodified inputs before adversarial generation begins.

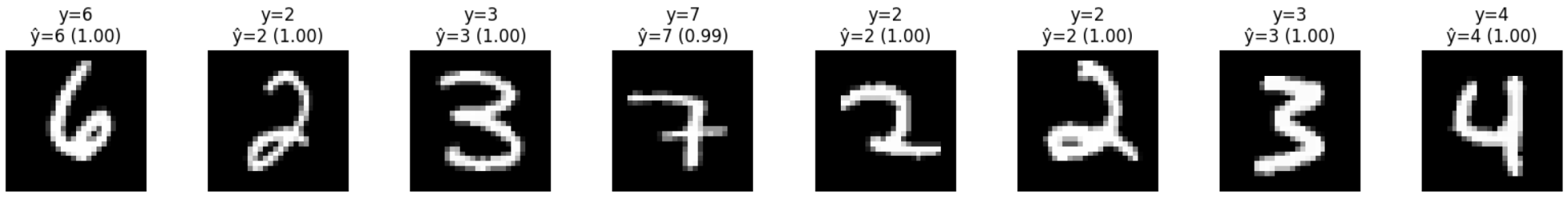

Figure 1: Representative clean MNIST samples used for sanity checking before adversarial perturbations were generated.

### 3.2 Experiment 2: Text Robustness Evaluation Using DistilBERT on SMS Spam

Experiment 2 evaluates robustness behavior in text classification using the SMS Spam Collection, a public corpus of 5,574 SMS messages labeled as ham or spam [9]. The dataset was loaded through the Hugging Face Datasets library [11], standardized into text/label fields, and split with a fixed seed into training, validation, and test partitions of 80%, 10%, and 10%, respectively, resulting in 4,459 training messages, 557 validation messages, and 558 test messages. This split was chosen to keep the experiment reproducible while preserving enough labeled data for both fine-tuning and held-out evaluation.

The text classifier was built using distilbert-base-uncased as a binary sequence-classification model. DistilBERT was chosen because it preserves much of BERT's contextual language-understanding capacity while reducing model size and computational cost, making it suitable for small-scale and education-oriented experiments [7], [8]. Tokenization was performed with the corresponding pretrained tokenizer, and dynamic padding was handled through a data collator. Fine-tuning was performed with the Hugging Face Transformers framework [10] using a learning rate of $2 \times 10^{-5}$, a per-device training batch size of 16, a per-device evaluation batch size of 32, one training epoch, weight decay of 0.01, and model selection based on validation F1-score.

Unlike the image experiment, the text component was intentionally scoped as a controlled robustness evaluation rather than an automated adversarial attack study. For a small set of spam messages from the test set, the experiment applies a short sequence of predefined, non-optimized perturbations and records the classifier's spam probability after each step. The perturbation sequence includes mild character substitutions, added whitespace noise, and the appending of a short benign filler phrase. These transformations were selected to preserve an educational and ethics-aware scope while still allowing changes in model confidence to be measured under altered input conditions.

To make the perturbation design concrete, representative spam-style examples drawn from the SMS Spam Collection are described here. For instance, one dataset message begins "WINNER!! As a valued network customer … claim code KL341 …", and another begins "URGENT! You have won a 1 week FREE membership …". In the current pipeline, such messages are not semantically rewritten or optimized against the classifier; instead, they are subjected to limited surface-level changes that preserve the overall spam intent [9], [11].

**Example A:** A message starting with *"WINNER!! As a valued network customer ..."* can be perturbed to forms such as *"W1NNER!! As a valued netw0rk cust0mer ..."*, then with inserted spacing noise, and finally with a short benign suffix such as *"Thanks for your time."* appended at the end.

**Example B:** A message beginning *"URGENT! You have won a 1 week FREE membership ..."* can be changed to *"URG3NT! Y0u have won a 1 week FREE membership ..."*, followed by spacing noise and the same non-optimized filler phrase.

These two examples align conceptually with surface-level perturbation strategies discussed in prior adversarial-NLP work, although the present study does not implement automated attack search or optimized token replacement [13]–[15].

### 3.3 Evaluation Protocol and Reproducibility

Across both experiments, the evaluation protocol separates baseline performance from perturbed-input behavior. In the MNIST experiment, the primary outputs are clean accuracy, attacked accuracy under each ε value, and qualitative clean/adversarial/perturbation panels. In the text experiment, the outputs include clean-set accuracy, precision, recall, and F1-score, together with a confusion matrix and per-example spam-probability trajectories before and after normalization. These metrics are standard for supervised classification [12].

Both experiments fix the random seed to 42 and save figures and CSV tables automatically into dedicated results directories. The ethical scope of the two experiments also differs by design: the MNIST experiment demonstrates successful white-box evasion on a standard benchmark, whereas the text experiment remains a local, controlled robustness demonstration on a public spam dataset and includes no automated search or deployment guidance for bypassing real-world filters.

## 4. Results

This section reports the outcomes of the two experiments introduced in the previous section. The results are presented separately for the image-based MNIST study and the text-based DistilBERT spam-classification study, followed by a brief cross-experiment summary. In line with standard classification practice, the reported metrics emphasize both overall correctness and class-sensitive behavior rather than relying on a single score alone [5], [6], [12].

### 4.1 Results of the MNIST Image Experiment

The MNIST classifier achieved a clean test accuracy of 98.63%, indicating that the compact convolutional model learned the digit-recognition task effectively before any perturbation was introduced. This high clean-data baseline is important because it shows that later performance degradation is attributable to the adversarial perturbations rather than to a weak classifier [1]–[3].

| ε | FGSM accuracy (%) | PGD accuracy (%) |
|---|---|---|
| 0.00 | 98.63 | 98.63 |
| 0.02 | 97.67 | 97.58 |
| 0.05 | 95.03 | 93.90 |
| 0.10 | 84.62 | 75.14 |
| 0.15 | 60.20 | 32.47 |
| 0.20 | 27.89 | 6.02 |
| 0.30 | 1.72 | 0.41 |

Table 2: MNIST clean and attacked test accuracy under FGSM and PGD across the evaluated perturbation budgets.

Under FGSM, accuracy decreased steadily as the perturbation budget increased. Accuracy remained high at 97.67% for ε = 0.02, fell to 95.03% at ε = 0.05, and dropped more sharply to 84.62% at ε = 0.10. The decline became much more severe at larger perturbation levels, reaching 60.20% at ε = 0.15, 27.89% at ε = 0.20, and 1.72% at ε = 0.30. This monotonic reduction confirms that even a single gradient-based perturbation step can significantly disrupt classification when the attack budget becomes sufficiently large [2].

The PGD results were even more damaging than the FGSM results across most perturbation levels, which is consistent with the role of PGD as a stronger iterative first-order attack [3]. At ε = 0.02, PGD accuracy was 97.58%, which is nearly identical to FGSM. However, the gap widened as ε increased: accuracy was 93.90% at ε = 0.05, 75.14% at ε = 0.10, 32.47% at ε = 0.15, 6.02% at ε = 0.20, and 0.41% at ε = 0.30. These results show that the iterative attack produced substantially greater degradation than FGSM under the same perturbation budget, especially in the middle and upper parts of the ε range [3].

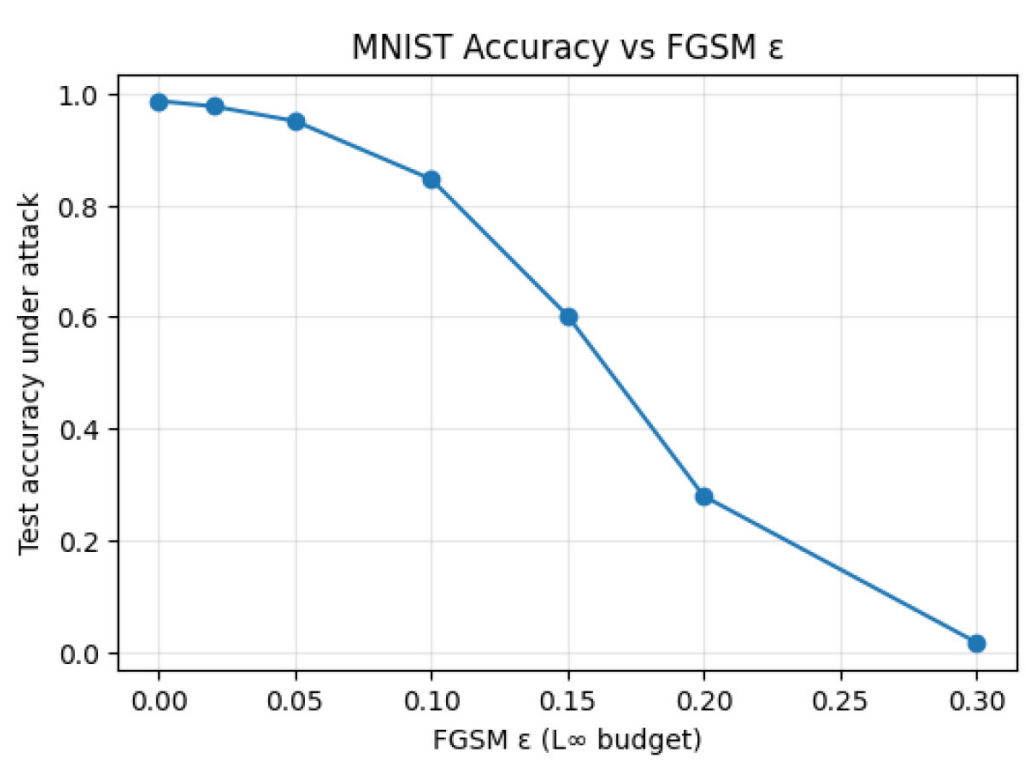

Figure 2: MNIST test accuracy under FGSM as the perturbation budget ε increases.

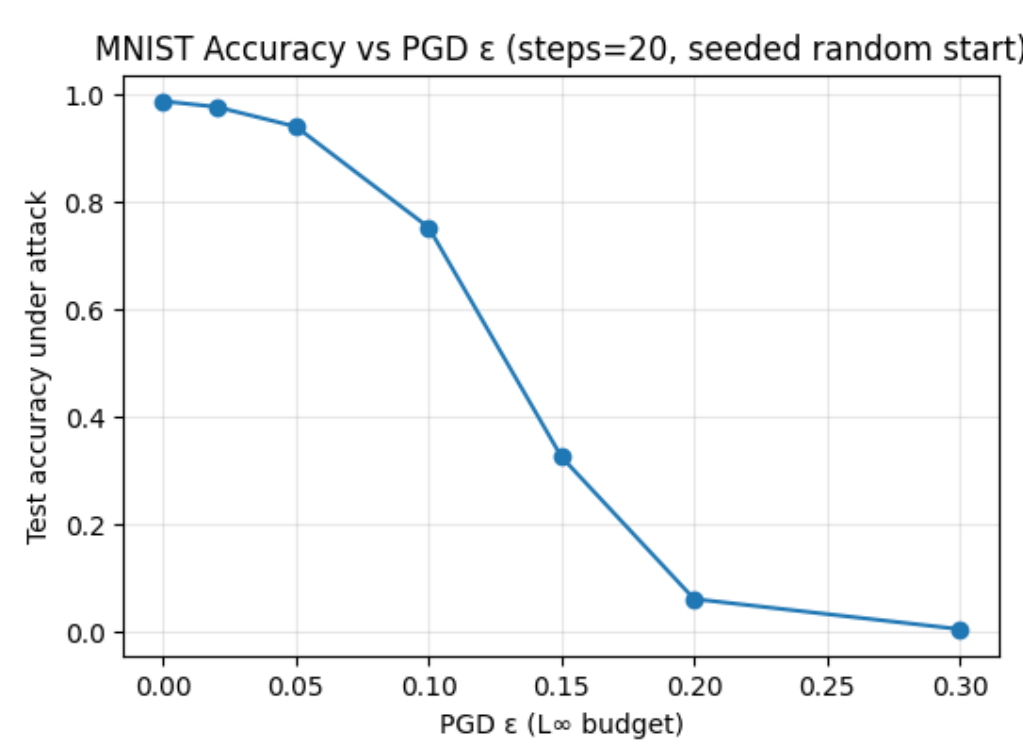

Figure 3: MNIST test accuracy under PGD (20 steps) across the same ε values used for FGSM.

The qualitative example provides an especially clear illustration of successful evasion. In the clean condition, the selected image was correctly predicted as digit 7 with confidence 0.9986. After FGSM perturbation at ε = 0.15, the same image was classified as digit 1 with confidence 0.7069. This example is useful because it shows not only a label flip, but also a wrong prediction made with non-trivial confidence. In other words, the model was not merely uncertain; it was confidently wrong after perturbation, which is a defining characteristic of adversarial vulnerability [2], [3].

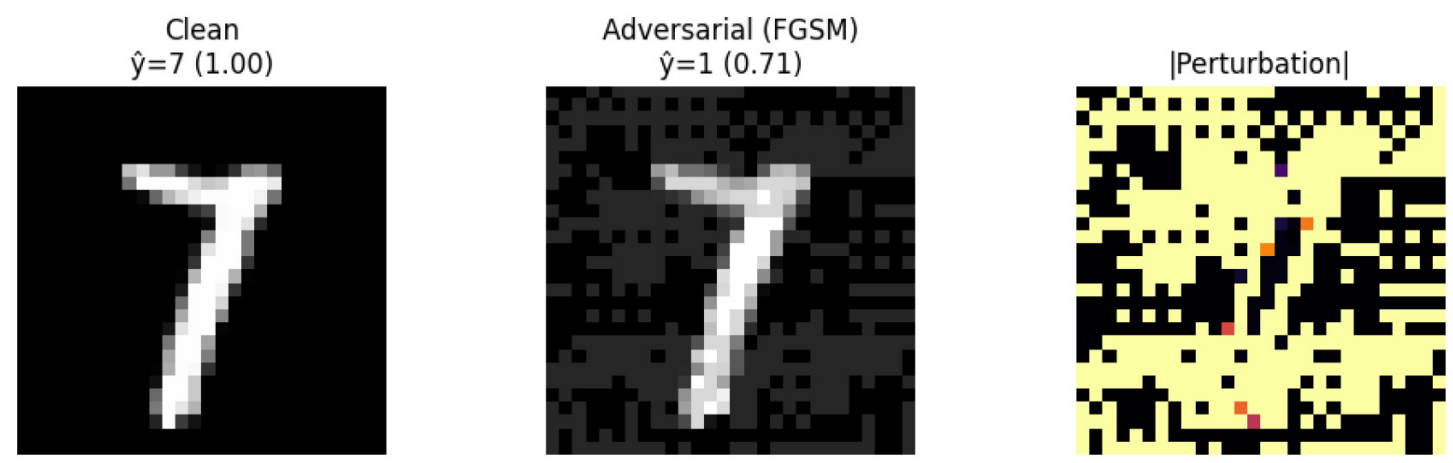

Figure 4: Representative FGSM example showing a clean image, its adversarial counterpart, and the absolute perturbation map.

The simple preprocessing defense produced only a partial improvement. For the PGD adversarial inputs at ε = 0.15 from Table 2, the bit-depth reduction step increased accuracy from 32.47% to 46.56%. Across five random starts (seeds 42 to 46), PGD accuracy at ε = 0.15 ranged from 32.36% to 32.54% (mean 32.44%, standard deviation 0.07 points), and the defense gained between 14.08 and 14.60 points, so its effect is far larger than the run-to-run variation. Although this increase is meaningful, the defended performance remained far below the clean baseline of 98.63%. This pattern supports a cautious interpretation: lightweight preprocessing may mitigate some perturbation effects, but it does not eliminate the robustness problem. Such caution is consistent with prior work warning that apparently successful defenses must be interpreted carefully under stronger evaluation [4].

## 4.2 Results of the Text DistilBERT Experiment

The DistilBERT text classifier also achieved strong clean-data performance. On the held-out test split, the model obtained an accuracy of 98.75%, precision of 97.06%, recall of 92.96%, and F1-score of 94.96%. These values indicate that the classifier learned the ham-versus-spam distinction effectively under standard evaluation, and they are in line with the strong baseline behavior commonly associated with transformer-based text classifiers [6]–[8], [12].

| Metric | Value (%) |
|---|---|
| Accuracy | 98.75 |
| Precision | 97.06 |
| Recall | 92.96 |
| F1-score | 94.96 |

Table 3: Clean test metrics for the DistilBERT SMS spam classifier.

The confusion-matrix output further supports this result by showing that most test messages were classified correctly, with 485 ham messages and 66 spam messages placed in their correct classes, alongside only 2 ham-to-spam errors and 5 spam-to-ham errors. From a reporting perspective, this is important because spam detection should not be evaluated by accuracy alone; precision, recall, and F1-score are all relevant to understanding the trade-off between blocking malicious messages and preserving legitimate ones [5], [6], [12].

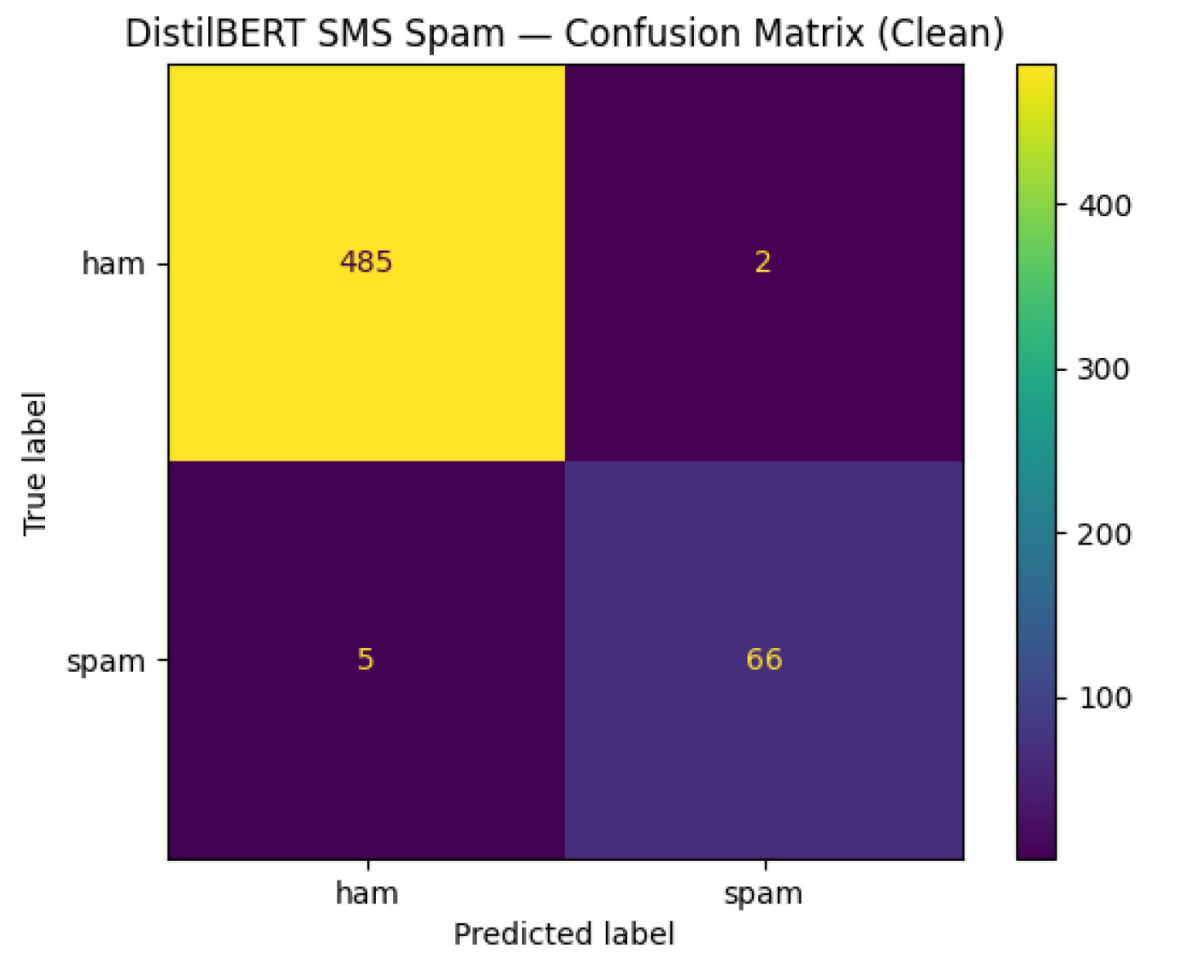

Figure 5: Confusion matrix for the clean DistilBERT SMS spam classifier on the held-out test split.

Unlike the image experiment, however, the text perturbation study did not produce a uniformly strong evasion outcome under the current perturbation design. For most displayed examples, the predicted spam probability remained high throughout the perturbation sequence, indicating that the selected perturbations caused only small confidence changes and usually did not trigger a label flip. At the same time, the trajectory plot also shows that sensitivity can vary across examples. Example 5 is the exception: the clean model assigned it a spam probability near 0, so this spam message was missed before any perturbation, and the character substitutions raised its spam probability to near 1. For that message the perturbation made the spam more detectable rather than less, the opposite of an evasion effect.

The normalization experiment produced modest but still informative changes. Several probability traces shifted downward after normalization, showing that the preprocessing step had a measurable effect on confidence; however, the affected examples generally remained on the spam side of the decision boundary. Example 5 is again the exception: with normalization applied, its spam probability stayed near 0 at every step, so the message that the substitutions had made detectable went back to being missed. Accordingly, the current text pipeline is best interpreted as a controlled robustness or sensitivity evaluation rather than as a strong text-evasion demonstration [6], [8], [12].

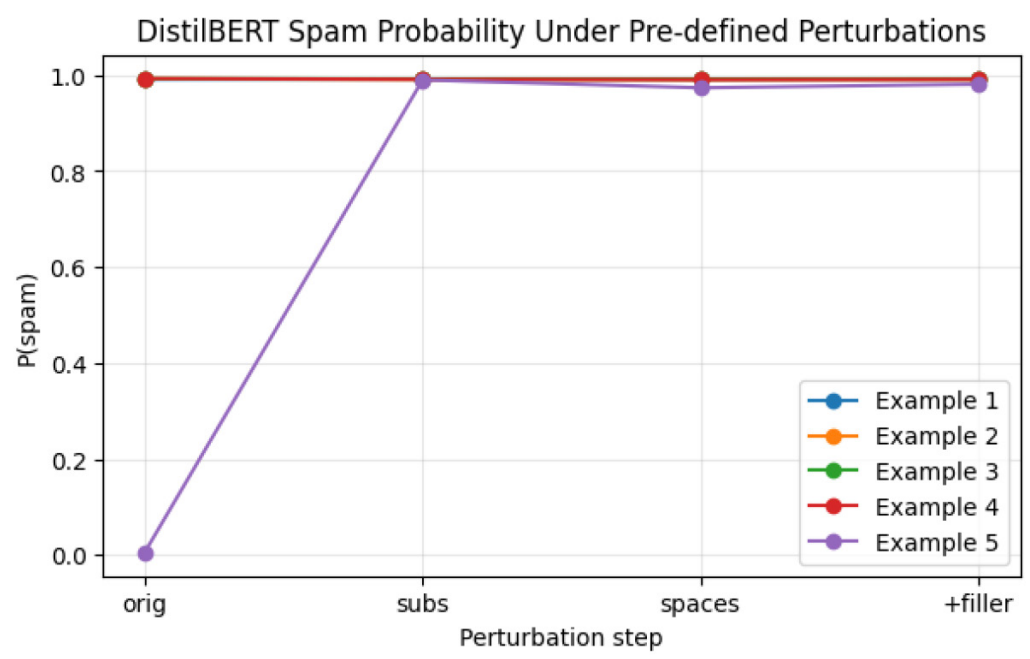

Figure 6: Spam-probability trajectories for selected messages under the pre-defined text perturbation sequence.

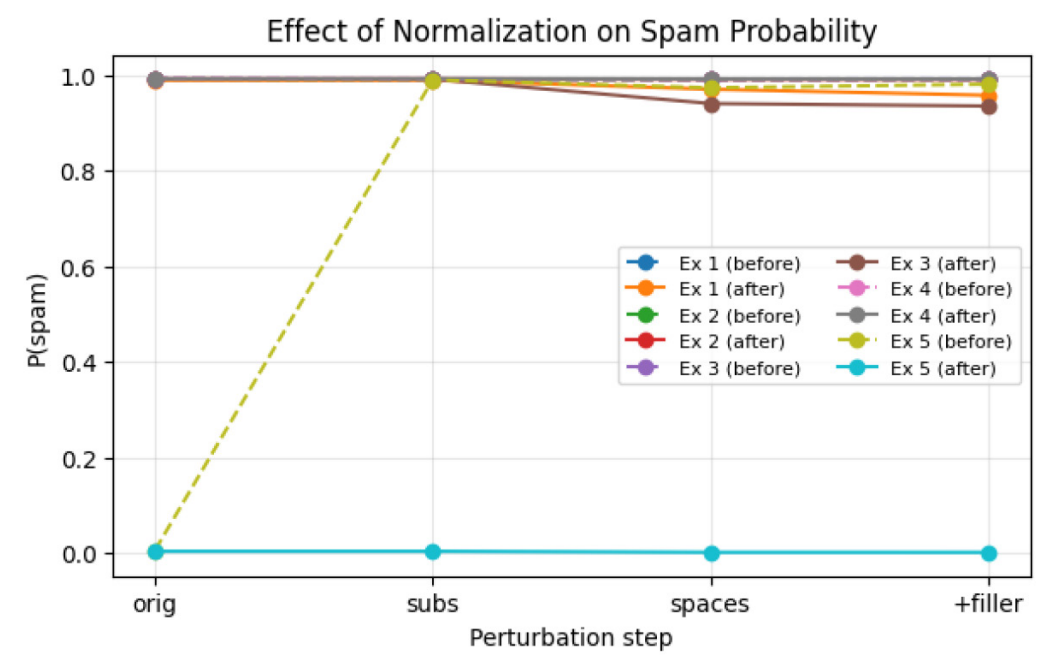

Figure 7: Effect of the normalization step on the spam-probability trajectories of the selected messages.

### 4.3 Cross-Experiment Summary

The combined results reveal a modality-dependent pattern. In the MNIST benchmark, small gradient-based perturbations produced large and systematic accuracy losses, particularly under the stronger PGD attack, whereas in the DistilBERT spam experiment the selected surface-level perturbations caused only modest probability shifts in the displayed examples and no flip from spam to ham. The image experiment is therefore a clear benchmark demonstration of successful evasion, while the text experiment shows that a compact transformer classifier in a local spam-detection setting can remain comparatively stable under mild perturbations even though its confidence shifts [2]–[4], [13]–[15].

## 5. Discussion

### 5.1 Interpretation of the MNIST Findings

The MNIST results are consistent with the expected behavior of first-order white-box attacks. FGSM reduced accuracy progressively as the perturbation budget increased, while PGD produced even larger losses under the same budget at most values of ε. This outcome is theoretically and empirically aligned with prior adversarial-learning literature, which treats PGD as a stronger iterative baseline than single-step attacks such as FGSM because it repeatedly follows the loss gradient and re-projects the sample into the allowable perturbation region [2], [3].

The bit-depth reduction experiment should also be interpreted cautiously. Although the defense increased PGD accuracy at $\varepsilon = 0.15$, the defended performance remained far below the clean baseline. This partial recovery is informative from a teaching perspective because it shows that simple preprocessing may mitigate some perturbation effects, but it does not justify a claim of robustness. This caution is supported by prior work showing that many defenses that appear effective under limited evaluation can fail under stronger or adaptive adversarial testing [4].

## 5.2 Interpretation of the Text Findings

The text experiment leads to a more nuanced interpretation. On clean data, DistilBERT achieved strong performance, which is consistent with prior work showing that transformer-based classifiers often establish strong baselines in text categorization tasks [6]–[8]. However, the controlled perturbation study did not yield a comparably strong bypass result under the current perturbation set. Instead, the spam probabilities generally remained high after mild substitutions, whitespace changes, and the addition of a short benign phrase.

This outcome does not imply that transformer-based spam classifiers are broadly robust to all adversarial manipulation. Rather, it indicates that the particular local model, dataset, and perturbation design used in this study were not sufficient to produce major confidence collapse or frequent label flipping in the displayed cases. That interpretation is compatible with prior adversarial-NLP research, which shows that text attacks are often constrained by semantic preservation, tokenization behavior, and the discrete structure of language, making the design of effective perturbations more complex than in continuous image spaces [13]–[15].

## 5.3 Comparative Summary

Table 4 summarizes the main findings of the two experiments and highlights the different conclusions that can be drawn from each modality.

| Criterion | MNIST image experiment | DistilBERT spam experiment |
|---|---|---|
| Clean baseline | 98.63% accuracy on the test set. | 98.75% accuracy, 97.06% precision, 92.96% recall, and 94.96% F1-score on the test set. |
| Perturbation mechanism | White-box gradient-based attacks (FGSM and PGD) under an $L\infty$ $\varepsilon$ budget. | Controlled, pre-defined text perturbations with optional normalization. |
| Observed effect | Large and systematic accuracy degradation; PGD stronger than FGSM at most $\varepsilon$ values. | Only modest probability shifts in the displayed examples; no flip from spam to ham under the current perturbation set. |
| Main interpretation | Clear successful evasion demonstration under benchmark image conditions [2], [3]. | Local robustness/sensitivity study rather than a strong evasion demonstration [14], [15]. |

Table 4: Comparative summary of the results.

## 5.4 Limitations

Several limitations should be acknowledged. First, MNIST is a simplified benchmark and does not represent the full visual complexity of modern image-recognition tasks [1]. Second, the text experiment uses a relatively small SMS dataset with short messages and a limited set of manually specified perturbations, which restricts the generality of the text-side conclusions [9]. Third, the text pipeline was intentionally constrained to an ethics-aware, educational scope and therefore did not include automated search or optimization procedures of the kind commonly explored in broader adversarial-NLP frameworks [14], [15]. These limitations do not invalidate the current findings, but they do define the boundaries within which the results should be interpreted.

### 5.5 Next Steps

The most important next step is to strengthen the text evaluation while preserving the project's responsible scope. A stronger second-round study could evaluate a larger set of messages, report aggregate probability-drop statistics, count label flips, and compare the current transformer baseline with simpler text baselines or alternative preprocessing choices. Such an extension would better balance the image and text components.

## 6. Conclusion

This paper presented a two-part, reproducible study of adversarial robustness spanning image classification and text classification. The MNIST experiment demonstrated a clear evasion effect: both FGSM and PGD reduced classifier accuracy as the perturbation budget increased, and PGD consistently produced stronger degradation than FGSM under the same attack constraints. The text experiment showed a different pattern. DistilBERT achieved strong clean-data performance on SMS spam detection, but the current set of pre-defined perturbations caused only limited confidence changes and did not provide equally strong evidence of successful bypass in the displayed cases.

Taken together, these results support two conclusions. First, high clean-data performance should not be treated as evidence of robustness, especially in image settings where first-order attacks can produce severe degradation. Second, robustness behavior is modality-dependent and must be established empirically rather than assumed; in text classification, the effect of perturbations depends heavily on how those perturbations interact with tokenization, semantics, and the attack design itself.

Future work should broaden the text-side perturbation analysis to many messages with summary statistics and examine how preprocessing and model choice affect the observed robustness.